\documentclass[12pt]{iopart}

\expandafter\let\csname equation*\endcsname\relax

\expandafter\let\csname endequation*\endcsname\relax

\usepackage{amsmath}
\usepackage{amsfonts}
\usepackage{iopams}
\usepackage{cite}
\usepackage[colorlinks=true, citecolor=blue]{hyperref}
\usepackage{graphicx}
\usepackage{hyperref}
\usepackage{graphics}
\usepackage{amssymb}
\usepackage{color}
\usepackage{bm}
\usepackage{float}
\usepackage[utf8x]{inputenc}

\begin{document}

\title[Entropic uncertainty lower bound for a two-qubit system coupled to a spin chain]{Entropic uncertainty lower bound for a two-qubit system coupled to a spin chain with Dzyaloshinskii-Moriya interaction}

\author{Soroush Haseli $^{1}$ \footnote{
Corresponding author}, Saeed Haddadi $^{2}$, and Mohammad Reza Pourkarimi $^{3}$}

\address{
$^{1}$ Faculty of Physics, Urmia University of Technology, Urmia, Iran\\
$^{2}$ Faculty of Physics, Semnan University, P.O.Box 35195-363, Semnan, Iran\\
$^{3}$ Department of Physics, Salman Farsi University of Kazerun, Kazerun, Iran\\
}

\ead{soroush.haseli@uut.ac.ir}
\ead{haddadi@semnan.ac.ir}
\ead{mrpourkarimy@gmail.com}

\vspace{10pt}
\begin{indented}
\item[]June 2020
\end{indented}

\begin{abstract}
The uncertainty principle is one of the key concepts in quantum theory. This principle states that it is not possible to measure two incompatible observables simultaneously and accurately. In quantum information theory, the uncertainty principle is formulated using the concept of entropy. In this work we consider the entropic uncertainty relation in the presence of quantum memory. We study the dynamics of entropic uncertainty bound for a two-qubit quantum system coupled to a spin chain with Dzyaloshinskii-Moriya interaction and we investigate the effect of environmental parameters on the entropic uncertainty bound. Notably, our results reveal that there exist some environmental parameters which can be changed to suppress the entropic uncertainty bound.

\end{abstract}

%

%
%
%
%

\section{Introduction}\label{intro}
In quantum theory, the uncertainty principle has been known as a key principle that determines the distinction between the classical and quantum world. The first description of the uncertainty principle was formally presented by Heisenberg in 1927 \cite{1}. Heisenberg's uncertainty relationship indicates that it is not possible to measure the location and momentum of a particle with high-precision simultaneously. The uncertainty principle for the two preferred observables $Q$ and $R$ was formulated by Robertson \cite{2} and Schr\"{o}dinger \cite{3} using the standard deviation as
\begin{equation}
\Delta Q \Delta R \geq \frac{1}{2}|\langle[Q, R]\rangle|,
\end{equation}
where $\Delta X=\sqrt{\left\langle X^{2}\right\rangle-\langle X\rangle^{2}}$ is the standard deviation of the $X \in \lbrace Q, R\rbrace$, $\langle X \rangle$ is the expectation value of $X$ and $[Q, R]=Q R-R Q$. In quantum information theory, information measurement criteria can be used to formulate the uncertainty relation. One of the criteria for measuring the amount of information is entropy. Moreover, it has been shown that the most appropriate quantity to define the uncertainty relations is entropy. The first entropic uncertainty relation (EUR) was formulated by using Shannon's entropy. The relation was proposed by Deutsch \cite{4}, then developed by Kraus\cite{5}, and in the last proved by Maassen and Uffink \cite{6}. If $\rho$ be the state of quantum system, then it has been shown that for the two desired observables $Q$ and $R$ the following EUR is established as
\begin{equation}\label{mas}
H(Q)+H(R) \geq \log _{2} \frac{1}{c},
\end{equation}
where $H(Q)=\sum_{i} p_{i} \log _{2} p_{i}$ and $H(R)=\sum_{j} m_{j} \log _{2} m_{j}$ are the Shannon entropy, $p_i=\langle q_i | \rho | q_i \rangle$, $m_j=\langle r_j | \rho | r_j \rangle$ and $c=\max _{i, j}\left\{\left|\left\langle q_{i} | r_{j}\right\rangle\right|^{2}\right\}$ where $| q_i \rangle$ and $|r_j \rangle $ are the eigenvectors of $Q$ and $R$, respectively. The above uncertainty relation can be interpreted by an interesting one-particle uncertainty game. At first, the particle is prepared in a quantum state $\rho$ by Bob. Bob sends the particle to Alice. Alice and Bob  agree on the measurement of two observables, $Q$ and $R$. Then one of the two observables $Q$ and $R$ is measured by Alice on her particle and she sends her choice to Bob. The uncertainty of Bob about the Alice's measurement is bounded by Eq. (\ref{mas}). Besides, the uncertainty game can be expanded by considering the two particles. In the new game, Bob prepares the correlated state $\rho_{AB}$ for two-particle system. He sends part $A$ to Alice and part $B$ is kept by himself. Part $B$, which is available to Bob, is used as a quantum memory particle. In this structure, Bob can guess the result of Alice's measurement with less uncertainty. The first entropic uncertainty relation in the presence of a quantum memory (EUR-QM) was introduced by Berta et al. as \cite{7}
\begin{equation}\label{berta}
S(Q | B)+S(R | B) \geq \log _{2} \frac{1}{c}+S(A | B),
\end{equation}
where $S(Q | B)=S\left(\rho^{Q B}\right)-S\left(\rho^{B}\right)$ and $S(R | B)=S\left(\rho^{R B}\right)-S\left(\rho^{B}\right)$ are the conditional von-Neumann entropies of the post-measurement states
\begin{align}
&\rho^{Q B}=\sum_{i}(\left|q_{i}\right\rangle_{A}\left\langle q_{i}\right| \otimes \mathbb{I}_{B}) \rho^{A B}(\left|q_{i}\right\rangle_{A}\left\langle q_{i}\right| \otimes \mathbb{I}_{B}),\nonumber\\
&\rho^{R B}=\sum_{j}(\left|r_{j}\right\rangle_{A}\left\langle r_{j}\right| \otimes \mathbb{I}_{B}) \rho^{A B}(\left|r_{j}\right\rangle_{A}\left\langle r_{j}\right| \otimes \mathbb{I}_{B}),
\end{align}
and $S(A|B)=S(\rho^{AB})-S(\rho^{B})$ is the conditional von-Neumann entropy.

The study of EUR-QM has been the subject of many activities in this field \cite{8,9,11,12,13,14,15,16,17,18,19,20,21,22,23,24,25,26,27,28,29,30,31,32,33,34,35,36,37,38,39,40,41,42,43,44,45,46,47,48,49,50,51,52,53,54,55,56,57,58,59,60,61,62,63,64,65,66}.
In Ref.\cite{62}, Adabi et al. proposed a new bound for EUR-QM as
 \begin{equation}\label{adabi}
 S(Q | B)+S(R | B) \geq \log _{2} \frac{1}{c}+S(A | B)+\max \{0, \delta\},
 \end{equation}
 where
\begin{equation}
\delta=I(A ; B)-(I(Q ; B)+I(R ; B)),
\end{equation}
in which $I(X;B)=S(\rho^{B})-\sum_x p_x S(\rho_x^{B})$ with $X \in \lbrace Q, R\rbrace$ is known as Holevo quantity,  $p_x=tr_{AB}(\Pi_{x}^{A}\rho^{AB}\Pi_x^{A})$ is the probability of $x$-th outcome, and $\rho_x^{B}=tr_{A}(\Pi_{x}^{A}\rho^{AB}\Pi_x^{A})/p_x$  is the state of Bob after the measurement of $X$ by Alice. Hereinafter, we use abbreviation EUB (entropic uncertainty bound) for the right-hand side of inequality (\ref{adabi}). As can be seen from (\ref{adabi}), the EUB has the additional term when compare with Berta's EUR-QM (\ref{berta}). It has been shown that the EUB in Eq. (\ref{adabi}) is tighter than Berta's bound \cite{62}. The uncertainty relationship has a wide range of applications in quantum information such as entanglement detection \cite{67,68,69,70}, quantum cryptography \cite{71,72} and quantum key distribution\cite{73,74}.

Decoherence \cite{d1,d2,d3,d4,d5,d6,d7,d8,d9} is an unavoidable phenomenon related to open quantum systems, which occurs because of their interaction with the environment. This causes the decay of quantum correlations, which are an essential resource for quantum information processing. Herein, we study the dynamical behavior of EUB for two-qubit coupled to a spin chain with Dzyaloshinskii-Moriya (DM) interaction. Due to the interaction of the bipartite quantum system $\rho_{AB}$ with the environment, the quantum correlation between the two parts decreases, so it is expected that Bob's uncertainty about Alice's measurement outcome increases. In this work, we study the effect of environmental parameters on the dynamics of EUB. We observe that there exist some environmental parameters that can be changed to suppress the uncertainty bound. Briefly, the work is organized as follows. In Sec. \ref{sec2} we discuss about the model which is used in this work. In Sec. \ref{sec3} we analyze the EUB for a two-qubit system coupled to a spin chain with DM interaction. Finally, conclusions are presented in Sec. \ref{sec4}.

\section{The Model}\label{sec2}
Let's consider a model in which a two-qubit system interacts transversely with an environment that is a general XY spin chain
with $z$-component DM interaction. The total Hamiltonian for a $N$-qubit system which transversely coupled to an environmental spin chain is given by ($\hbar=1$) \cite{Yan}
\begin{equation}\label{Hamiltonian}
H=H_{E}^{(\lambda)}+H_{I},
\end{equation}
where
\begin{equation}\begin{aligned}
H_{E}^{(\lambda)}=&-\sum_{l}^{N}\left(\frac{1+\gamma}{2} \sigma_{l}^{x} \sigma_{l+1}^{x}+\frac{1-\gamma}{2} \sigma_{l}^{y} \sigma_{l+1}^{y}+\lambda \sigma_{l}^{z}\right) \\
&-\sum_{l}^{N} D\left(\sigma_{l}^{x} \sigma_{l+1}^{y}-\sigma_{l}^{y} \sigma_{l+1}^{x}\right),
\end{aligned}\end{equation}
and
\begin{equation}H_{I}=-g\left(\frac{1+\delta}{2} \sigma_{A}^{z}+\frac{1-\delta}{2} \sigma_{B}^{z}\right) \sum_{l}^{N} \sigma_{l}^{z},\end{equation}
are the self-Hamiltonian of the environment and interaction between system and environment, respectively. $\sigma_{l}^{\alpha}$ is the Pauli operator which acts on $l$-th sites of the spin chain. $\sigma^{z}_{AB}$ is the $z$-component of the Pauli operator on the two-qubit system. $\gamma$ quantifies the measure of anisotropy of exchange interaction in XY plane, $\lambda$ stands for the strength of transverse  magnetic field, $D$ shows the strength of DM interaction in the $z$-direction, and $g(\frac{1\pm \delta}{2})$ quantifies the coupling strength between qubits and the spin chain. By adjusting $\delta$ one can control the anisotropy of coupling strength of qubit with spin chain. $N$ stands for the number of sites for XY spin chain. The periodic boundary conditions are also considered as $\sigma^{\alpha}_{N+1}=\sigma^{\alpha}_{1}$. By considering the eigenstates of the operator $\frac{1+\delta}{2}\sigma_A^{z}+\frac{1-\delta}{2}\sigma_{B}^{z}$ as what follows
\begin{equation}
\left|\phi_{1}\right\rangle=|00\rangle_{1},\,\,\left|\phi_{2}\right\rangle=|01\rangle_{2},\,\,\left|\phi_{3}\right\rangle=|10\rangle_{3},\,\,\left|\phi_{4}\right\rangle=|11\rangle_{4},
\end{equation}
where $\vert 0 \rangle$ and $\vert 1 \rangle$ represents spin up and down respectively, one can rewrite the Hamiltonian in Eq. (\ref{Hamiltonian}) as
\begin{equation}
H=\sum_{\mu=1}^{4}\left|\phi_{\mu}\right\rangle\left\langle\phi_{\mu}\right| \otimes H_{E}^{\left(\lambda_{\mu}\right)},
\end{equation}
where $\lambda_\mu$'s are
\begin{equation}
\lambda_{1(4)}=\lambda \pm g, \,\, \lambda_{2(3)}=\lambda \pm g \delta,
\end{equation}
and $H_{E}^{(\lambda_{\mu})}$ can be obtain by substituting $\lambda$ with $\lambda_\mu$ in $H_{E}^{(\lambda)}$. To describe the evolution of a two-qubit system, it is necessary to identify the unitary  time-evolution operator $U(t)=\exp(-iHt)$. To achieve this, one can use Jordan-Wigner transformation to map the $H_{E}^{\lambda_{\mu}}$ on to a one dimensional spinless fermion system with creation and annihilation operators $c_{l}^{\dag}$ and $c_{l}$, respectively \cite{Sachdev}
\begin{align}
&\sigma_{l}^{x} =\prod_{s<l}\left(1-2 c_{s}^{\dagger} c_{s}\right)\left(c_{l}+c_{l}^{\dagger}\right),\nonumber\\
&\sigma_{l}^{y} =-i \prod_{s<l}\left(1-2 c_{s}^{\dagger} c_{s}\right)\left(c_{l}-c_{l}^{\dagger}\right),\nonumber\\
&\sigma_{l}^{z} =1-2 c_{l}^{\dagger} c_{l}.
\end{align}
The Hamiltonian will be diagonalized in the following way. At first, the Hamiltonian can be transformed into momentum space by Fourier transforms of a fermionic operator as $d_{k}=\frac{1}{\sqrt{N}} \sum_{l}^{N} c_{l} e^{-i 2 \pi l k / N}$ where $k=-M, \ldots, M$ and $M=(N-1) / 2$. In the next step, taking advantage of Bogoliubov transformation
\begin{equation}\begin{aligned}
&\eta_{k, \lambda_{\mu}}=\cos \frac{\theta_{k}^{\left(\lambda_{\mu}\right)}}{2} d_{k}-i \sin \frac{\theta_{k}^{\left(\lambda_{\mu}\right)}}{2} d_{-k}^{\dagger},
\end{aligned}\end{equation}
where
\begin{equation}\begin{aligned}
&\theta_{k}^{\left(\lambda_{u}\right)}=\arctan \left(\frac{\gamma \sin \frac{2 \pi k}{N}}{\lambda_{\mu}-\cos \frac{2 \pi k}{N}}\right),
\end{aligned}\end{equation}
one can diagonalize the Hamiltonian as
\begin{equation}
H_{E}^{\left(\lambda_{\mu}\right)}=\sum_{k} \Lambda_{k}^{\left(\lambda_{\mu}\right)}\left(\eta_{k, \lambda_{\mu}}^{\dagger} \eta_{k, \lambda_{\mu}}-\frac{1}{2}\right),
\end{equation}
where the spectrum $\Lambda_{k}^{\left(\lambda_{\mu}\right)}$ can be written as
\begin{equation}
\Lambda_{k}^{\left(\lambda_{\mu}\right)}=2\left(\varepsilon_{k}^{\left(\lambda_{\mu}\right)}+2 D \sin \frac{2 \pi k}{N}\right),
\end{equation}
with $\varepsilon_{k}^{\left(\lambda_{\mu}\right)}=\sqrt{\left(\lambda_{\mu}-\cos \frac{2 \pi k}{N}\right)^{2}+\gamma^{2} \sin ^{2} \frac{2 \pi k}{N}}$.

Assume that the initial state of the total system is in the product form as $\rho_{t o t}(0)=\rho_{A B}(0) \otimes \rho_{E}(0)$, in which $\rho_{AB}(0)$ and $\rho_E(0)$ are the initial state of the two-qubit system and environment, respectively. It is supposed that the initial state of the environment is pure i.e. $\rho_{E}(0)=\left|\psi_{E}(0)\right\rangle\left\langle\psi_{E}(0)\right|$. The whole system will evolve as a unitary process $\rho_{t o t}(t)=U(t) \rho_{t o t}(0) U^{\dagger}(t)$.  The evolution of the two-qubit system is achieved by taking partial trace over environment as
\begin{equation}\label{reduce}
\begin{aligned}
\rho_{A B}(t) &=\operatorname{Tr}_{E}\left[\rho_{t o t}(t)\right] \\
&=\sum_{\mu, v}^{4} F_{\mu \nu}(t)\left\langle\phi_{\mu}\left|\rho_{A B}(0)\right| \phi_{v}\right\rangle\left|\phi_{\mu}\right\rangle\left\langle\phi_{v}\right|,
\end{aligned}
\end{equation}
where $F_{\mu \nu}(t)=\left\langle\psi_{E}\left|U_{E}^{\dagger\left(\lambda_{\nu}\right)}(t) U_{E}^{\left(\lambda_{\mu}\right)}(t)\right| \psi_{E}\right\rangle$ and $U_{E}^{\left(\lambda_{\mu}\right)}(t)=\exp \left(-i H_{E}^{\left(\lambda_{\mu}\right)} t\right)$ are decoherence factor and time evolution operator respectively. Let us consider the conversion between the groundstate of the self-Hamiltonian $\vert G \rangle_{\lambda}$ and the ground state projected-Hamiltonian as \cite{Quan,Yuan}
\begin{equation}\label{trans}
|G\rangle_{\lambda}=\prod_{k>0}^{M}\left(\cos \Theta_{k}^{\left(\lambda_{\mu}\right)}+i \sin \Theta_{k}^{\left(\lambda_{\mu}\right)} \eta_{k, \lambda_{\mu}}^{\dagger} \eta_{-k, \lambda_{\mu}}^{\dagger}\right)|G\rangle_{\lambda_{\mu}},\end{equation}
with $\Theta^{\left(\lambda_{\mu}\right)}=\left(\theta_{k}^{\left(\lambda_{\mu}\right)}-\theta_{k}^{(\lambda)}\right) / 2$. By the help of Eq. (\ref{trans}) and performing long calculations, the decoherence factor can be obtained as follows \cite{Yan}
\begin{equation}\begin{aligned}
|F_{\mu \nu}(t)|=& \prod_{k>0}^{M}\left[1-\sin ^{2}\left(2 \Theta_{k}^{\left(\lambda_{\mu}\right)}\right) \sin ^{2}\left(\Lambda_{k}^{\left(\lambda_{\mu}\right)} t\right)-\sin ^{2}\left(2 \Theta_{k}^{\left(\lambda_{\nu}\right)}\right)\right.\\
& \times \sin ^{2}\left(\Lambda_{k}^{\left(\lambda_{\nu}\right)} t\right)+2 \sin \left(2 \Theta_{k}^{\left(\lambda_{\mu}\right)}\right) \sin \left(2 \Theta_{k}^{\left(\lambda_{\nu}\right)}\right) \\
& \times \sin \left(\Lambda_{k}^{\left(\lambda_{\mu}\right)} t\right) \sin \left(\Lambda_{k}^{\left(\lambda_{\nu}\right)} t\right) \cos \left(\Lambda_{k}^{\left(\lambda_{\mu}\right)} t-\Lambda_{k}^{\left(\lambda_{\nu}\right)} t\right)-4 \\
& \times \sin \left(2 \Theta_{k}^{\left(\lambda_{\mu}\right)}\right) \sin \left(2 \Theta_{k}^{\left(\lambda_{\nu}\right)}\right) \sin ^{2}\left(\Theta_{k}^{\left(\lambda_{\mu}\right)}-\Theta_{k}^{\left(\lambda_{\nu}\right)}\right) \\
&\left.\times \sin ^{2}\left(\Lambda_{k}^{\left(\lambda_{\mu}\right)} t\right) \sin ^{2}\left(\Lambda_{k}^{\left(\lambda_{\nu}\right)} t\right)\right]^{1 / 2}.
\end{aligned}\end{equation}
In the case that the decoherence factor tends to one, the two-qubit system is weakly affected by the spin chain environment. While for the situation in which the decoherence factor tends to zero, the two-qubit system is strongly influenced by the environment.

\section{Results}\label{sec3}
Let us consider the case in which Alice and Bob initially share the set of two-qubit states with the maximally mixed marginal states. This state can be
written as
\begin{equation}
\rho_{A B}(0)=\frac{1}{4}\left(\mathbb{I}_{A} \otimes \mathbb{I}_{B}+\sum_{i} r_{i} \sigma_{A}^{i} \otimes \sigma_{B}^{i}\right),\end{equation}
where $i$ takes values 1, 2 and 3 which represent the Pauli operators $x, y$, and $z$ respectively, and $r_{i}$ is the real parameter.
Let's assume that the system available to Alice and Bob is influenced by the spin chain environment.  According to Eq. (\ref{reduce}), the evolved density matrix in the computational basis $|00\rangle_{1}, |01\rangle_{2}, |10\rangle_{3}$ and $|11\rangle_{4}$ is obtained as follows
\begin{equation}\rho_{A B}(t)=\frac{1}{4}\left(\begin{array}{cccc}\label{edm}
1+r_{3} & 0 & 0 & \Gamma \\
0 & 1-r_{3} & \Omega & 0 \\
0 & \Omega^{*} & 1-r_{3} & 0 \\
\Gamma^{*} & 0 & 0 & 1+r_{3}
\end{array}\right),\end{equation}
where $\Gamma=\left(r_{1}-r_{2}\right) F_{14}(t)$ and $\Omega=\left(r_{1}+r_{2}\right) F_{23}(t)$. We consider the case in which Alice measures one of the two observables $Q=\sigma_x$ and $R=\sigma_z$. So, we get $c=1/2$ and the EUB (\ref{adabi}) for the evolved density matrix (\ref{edm}) is obtained as
\begin{equation}\label{eub1}
\textmd{EUB}\equiv1+S(A|B)+\max \lbrace 0,\delta \rbrace,
\end{equation}
where
\begin{eqnarray}
S(A|B)&=&-1-\frac{1-\Omega - r_3}{4}\log_2 \frac{1-\Omega - r_3}{4} - \frac{1+\Omega - r_3}{4}\log_2 \frac{1+\Omega - r_3}{4} \nonumber \\
&-&\frac{1-\Gamma + r_3}{4}\log_2 \frac{1-\Gamma + r_3}{4}-\frac{1+\Gamma + r_3}{4}\log_2  \frac{1+\Gamma + r_3}{4},
\end{eqnarray}
and
\begin{eqnarray}
\delta&=&-2+\frac{1-\Omega - r_3}{4}\log_2 \frac{1-\Omega - r_3}{4} + \frac{1+\Omega - r_3}{4}\log_2 \frac{1+\Omega - r_3}{4} \nonumber \\
&+&\frac{1-\Gamma + r_3}{4}\log_2 \frac{1-\Gamma + r_3}{4}+\frac{1+\Gamma + r_3}{4}\log_2  \frac{1+\Gamma + r_3}{4} \nonumber \\
&-&\frac{1-r_3}{2} \log_2 \frac{1-r_3}{4} - \frac{1+r_3}{2} \log_2 \frac{1+r_3}{4} \nonumber \\
&-&\frac{2-\Gamma -\Omega}{4} \log_2 \frac{2-\Gamma -\Omega}{8} - \frac{2 + \Gamma + \Omega}{4} \log_2 \frac{2 + \Gamma + \Omega}{8}. \nonumber \\
\end{eqnarray}
\\
\\

\begin{figure}[H]
  \centering
  \includegraphics[width=0.50\textwidth]{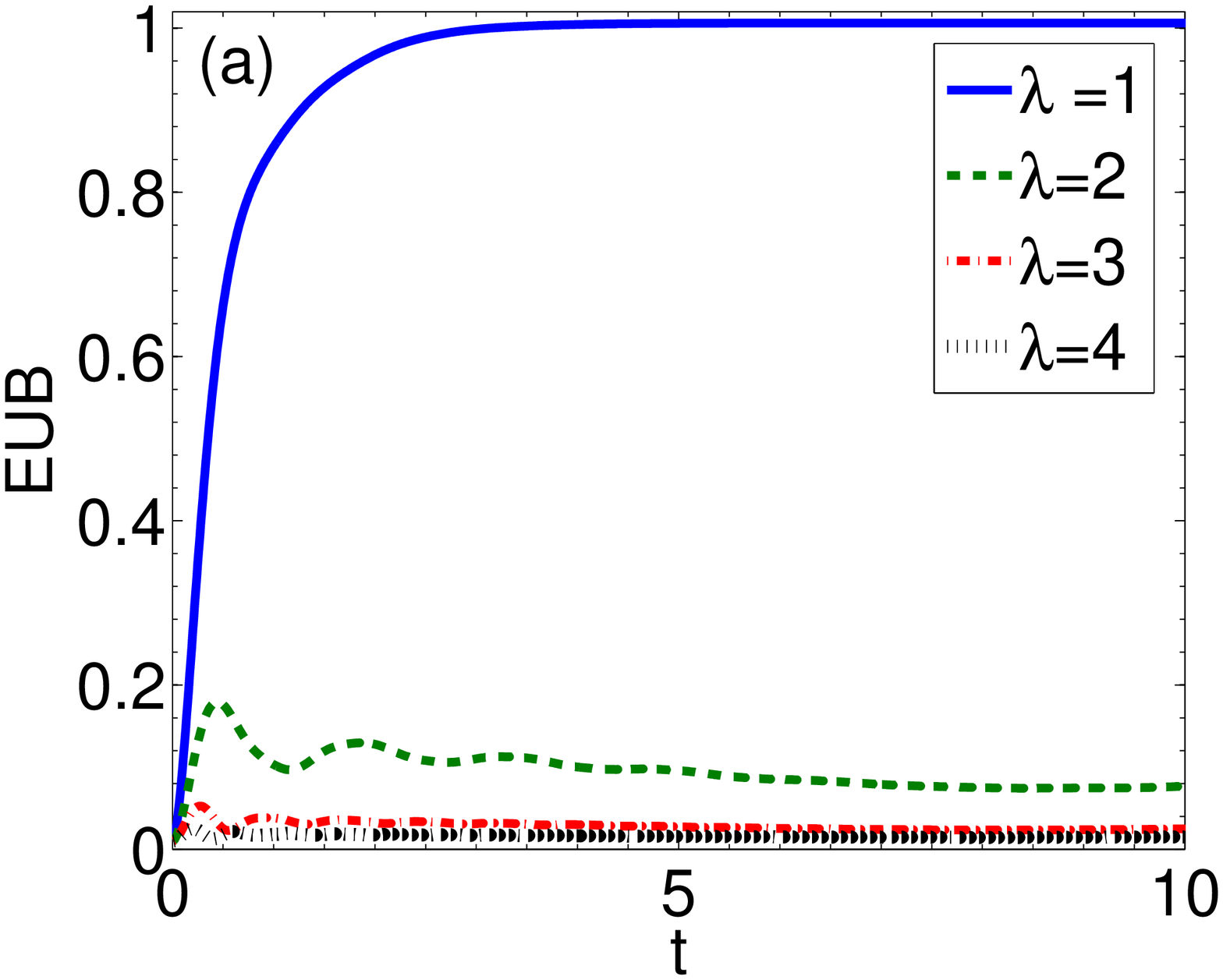}
  \includegraphics[width=0.48\textwidth]{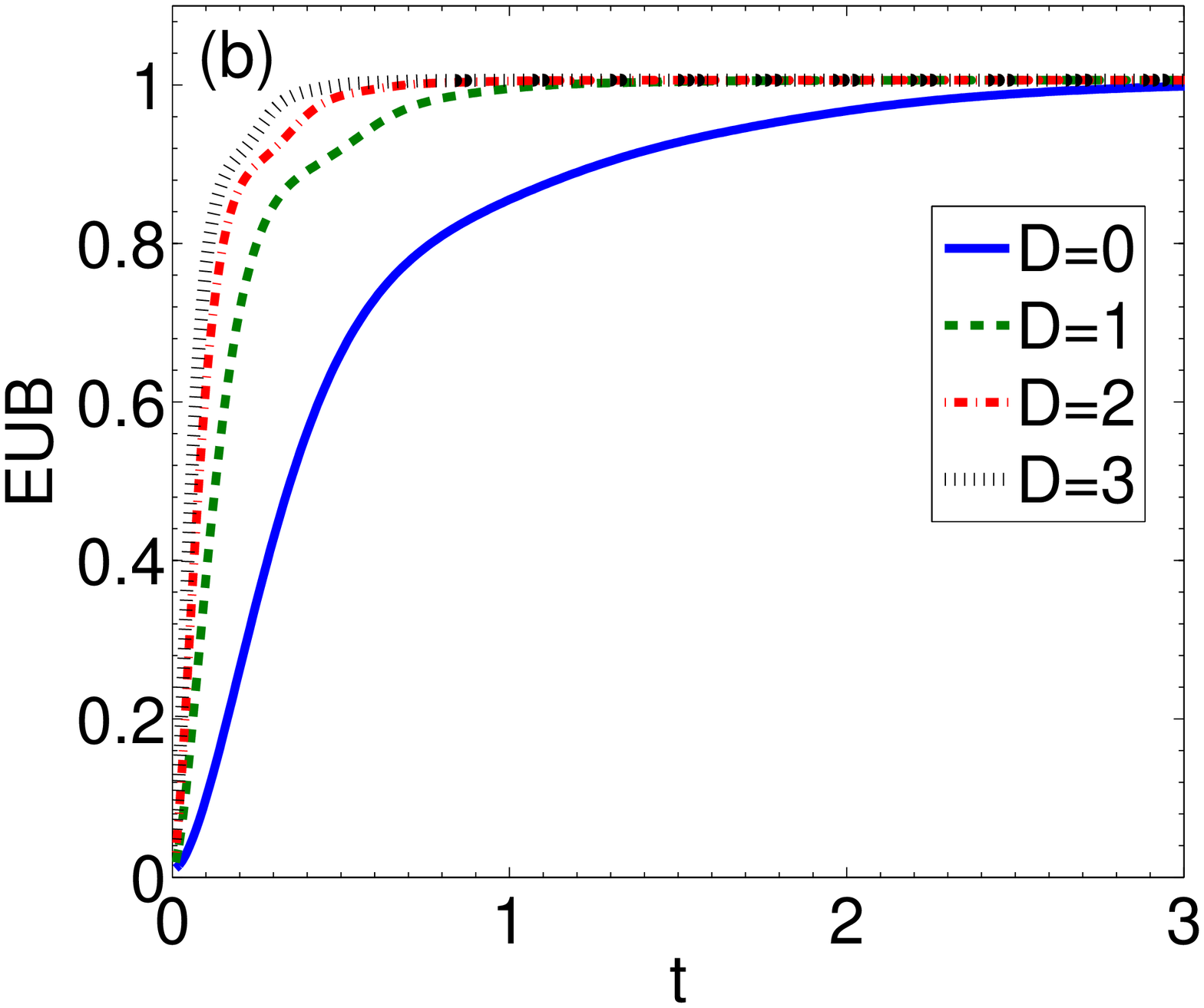}
  \includegraphics[width=0.49\textwidth]{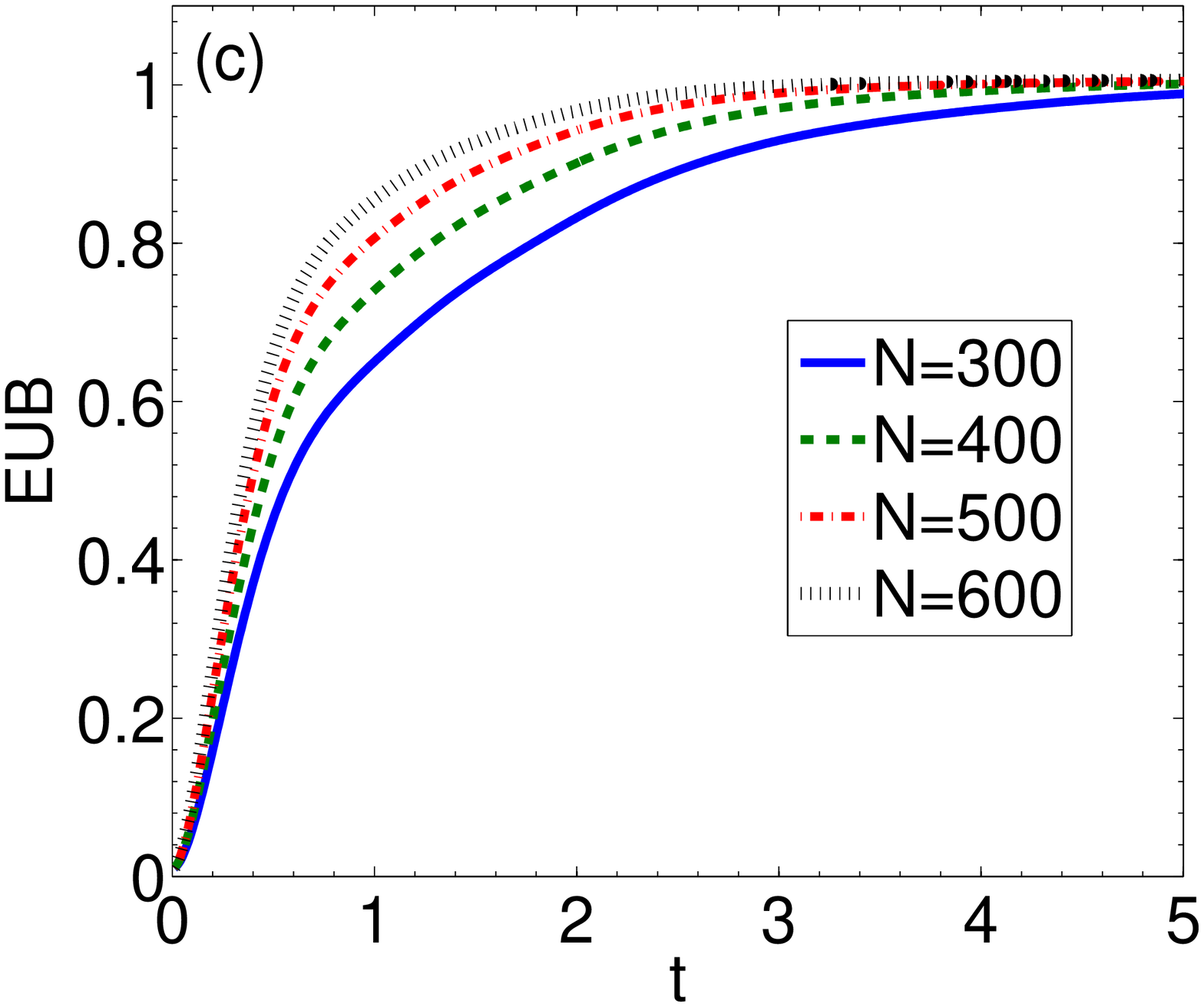}
  \includegraphics[width=0.49\textwidth]{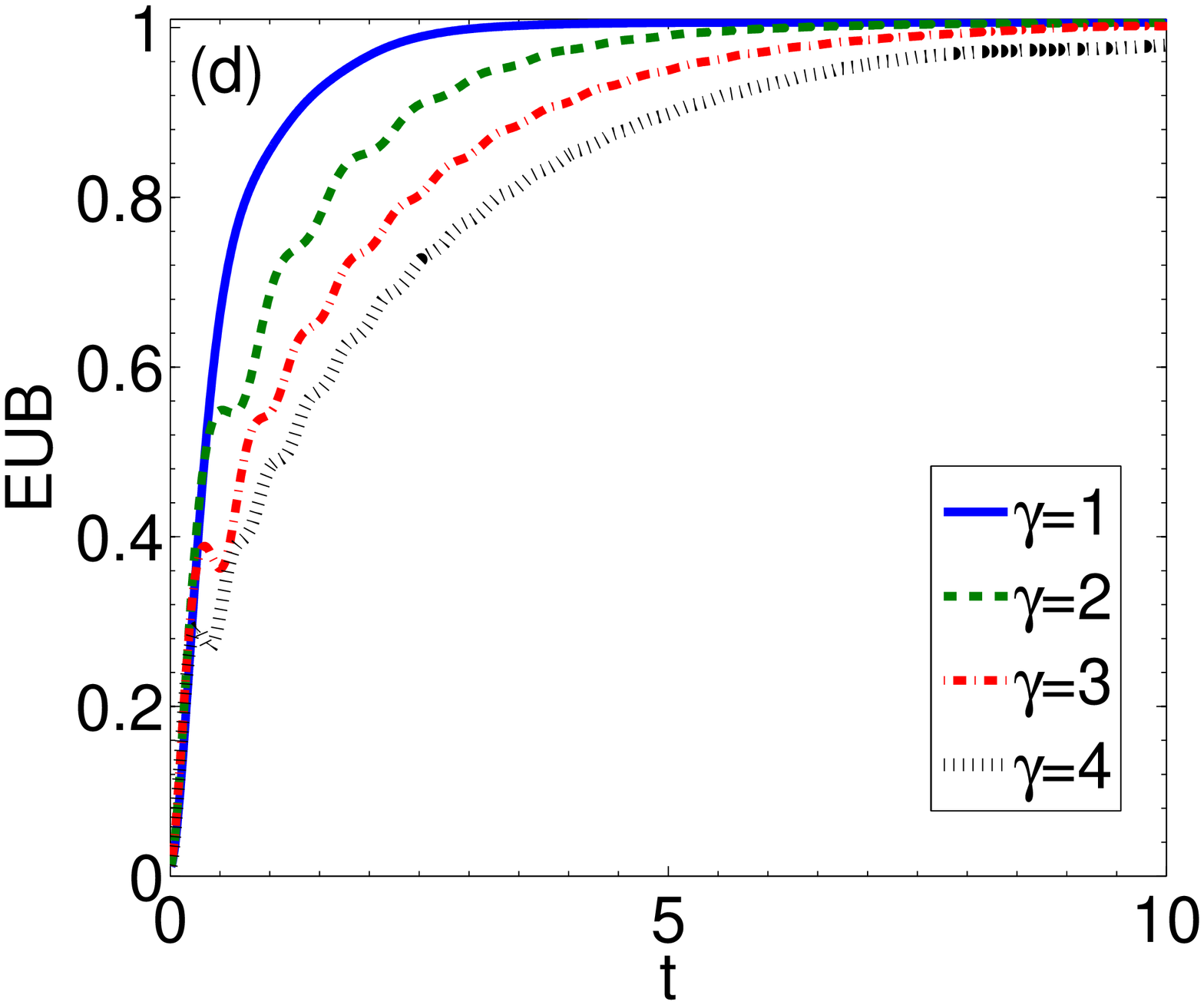}
\caption{The dynamics of entropic uncertainty bound for the case in which the initial state is pure with $r_1=1$ and $-r_2=r_3=1$. \textbf{(a)} For different values of the strength of transverse magnetic field $\lambda$, when $D=0$, $\gamma=1$, $\delta=0$, $g=0.05$ and $N=600$. \textbf{(b)} For different values of DM interaction $D$, when $\lambda=1$, $\gamma=1$, $\delta=0$, $g=0.05$ and $N=600$. \textbf{(c)} For different values of the spin number $N$, when $\lambda=1$, $\gamma=1$, $\delta=0$, $g=0.05$ and $D=0$. \textbf{(d)} For different values of the anisotropy $\gamma$, when $D=0$, $\lambda=1$, $\delta=0$, $g=0.05$ and $N=600$.}
\label{figure1}
\end{figure}

In Fig. \ref{figure1}, the EUB (\ref{eub1}) is plotted as a function of time for the case in which the initial state that is shared between Alice and Bob is pure with the state parameters $r_1=r_3=1$ and $r_2=-1,$ i.e. the initial state becomes a pure state as $\rho_{AB}(0)=|\Phi\rangle\langle\Phi|$ where $|\Phi\rangle=(|00\rangle+|11\rangle)/\sqrt{2}$ is a Bell state. Generally, it is clear that as a result of the interaction of the system with the environment, the quantum information of the system is lost. So when the system at the disposal of Alice and Bob interacts with the environment, Bob's uncertainty about the result of Alice's measurement will increase. However, the EUB can be controlled by adjusting the environmental parameter. Sometimes there are some environmental parameters that can be changed to reduce the uncertainty bound. Due to the fact that the initial state of the quantum system $\rho_{AB}$ is a maximally entangled pure state, the uncertainty at the initial moment of interaction is zero, and Bob can correctly guess the result of Alice's measurement at this time. When the interaction begins, it is observed that the uncertainty bound increases with the continuation of the interaction.

Fig. \ref{figure1}(a) shows the dynamics of EUB for different values of the transverse magnetic field strength $\lambda$. As can be seen, the EUB decreases with increasing parameter $\lambda$. In other words, the EUB can be suppressed by increasing the strength of the transverse magnetic field.
In Fig. \ref{figure1}(b), the dynamics of EUB is represented for different values of the strength of $z$-component DM interaction. Obviously, the EUB is enhanced by increasing the strength of DM interaction. Therefore, increasing this parameter increases Bob's uncertainty about the result of Alice's measurement.

\begin{figure}[H]
  \centering
  \includegraphics[width=0.49\textwidth]{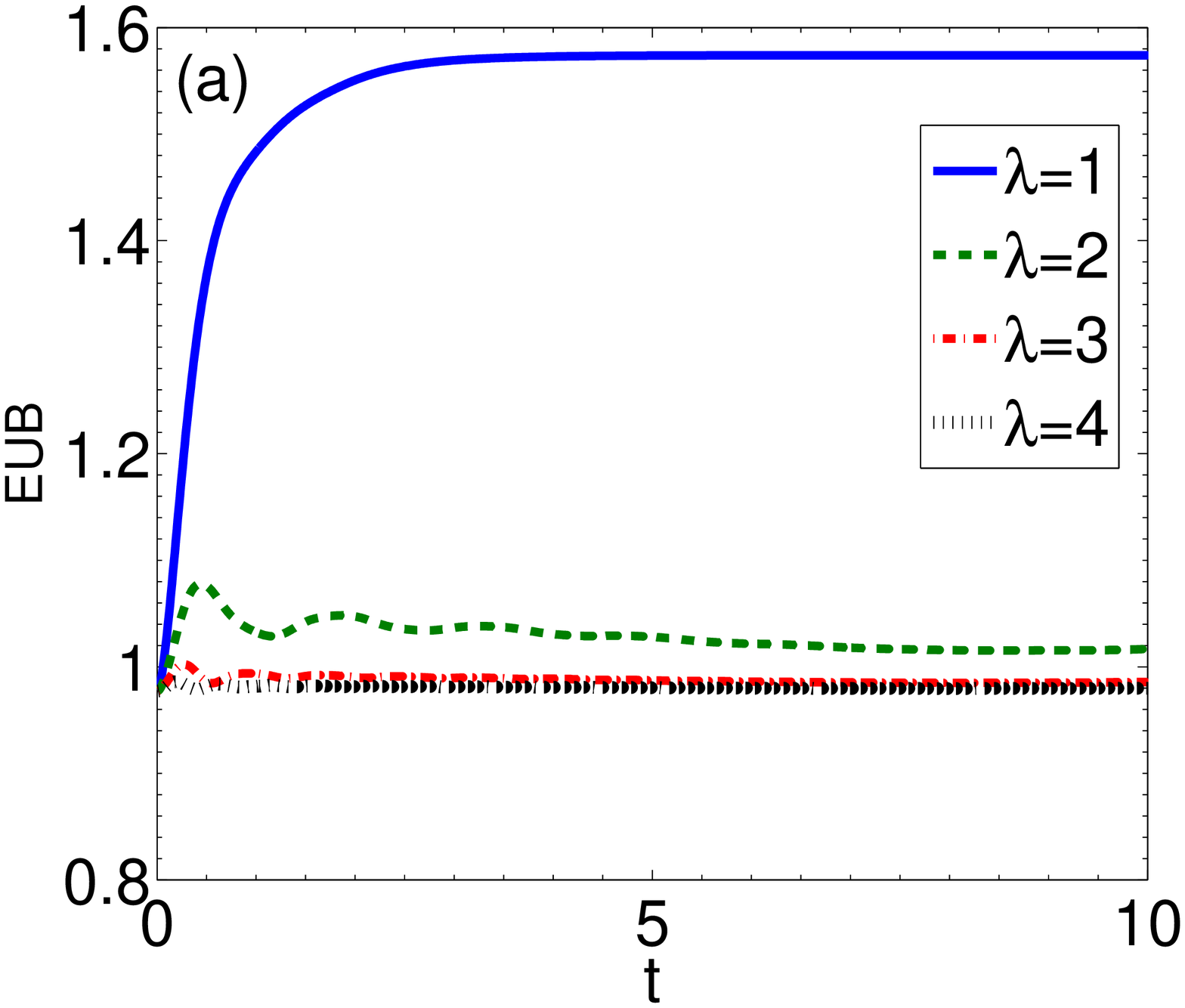}
  \includegraphics[width=0.49\textwidth]{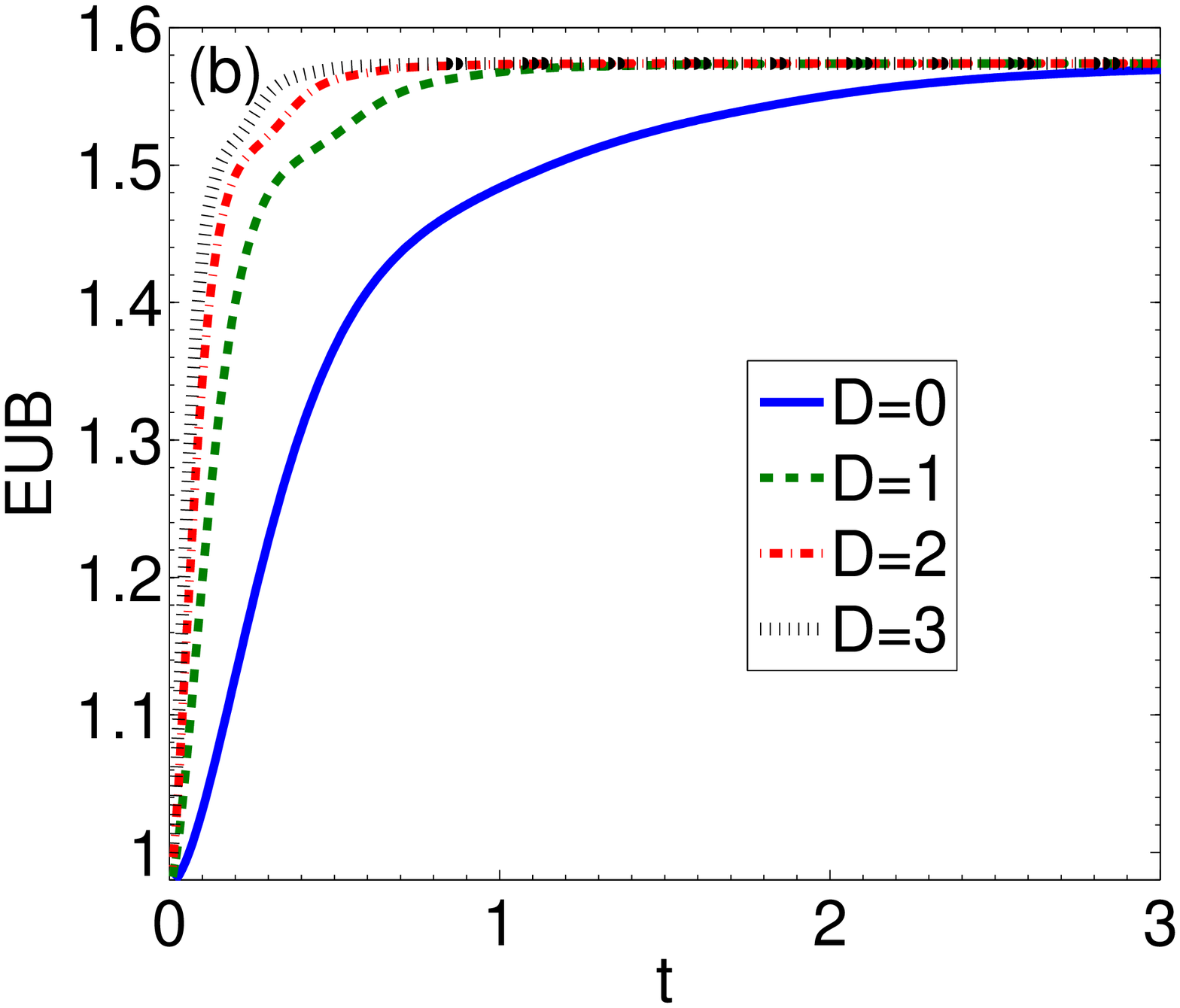}
  \includegraphics[width=0.49\textwidth]{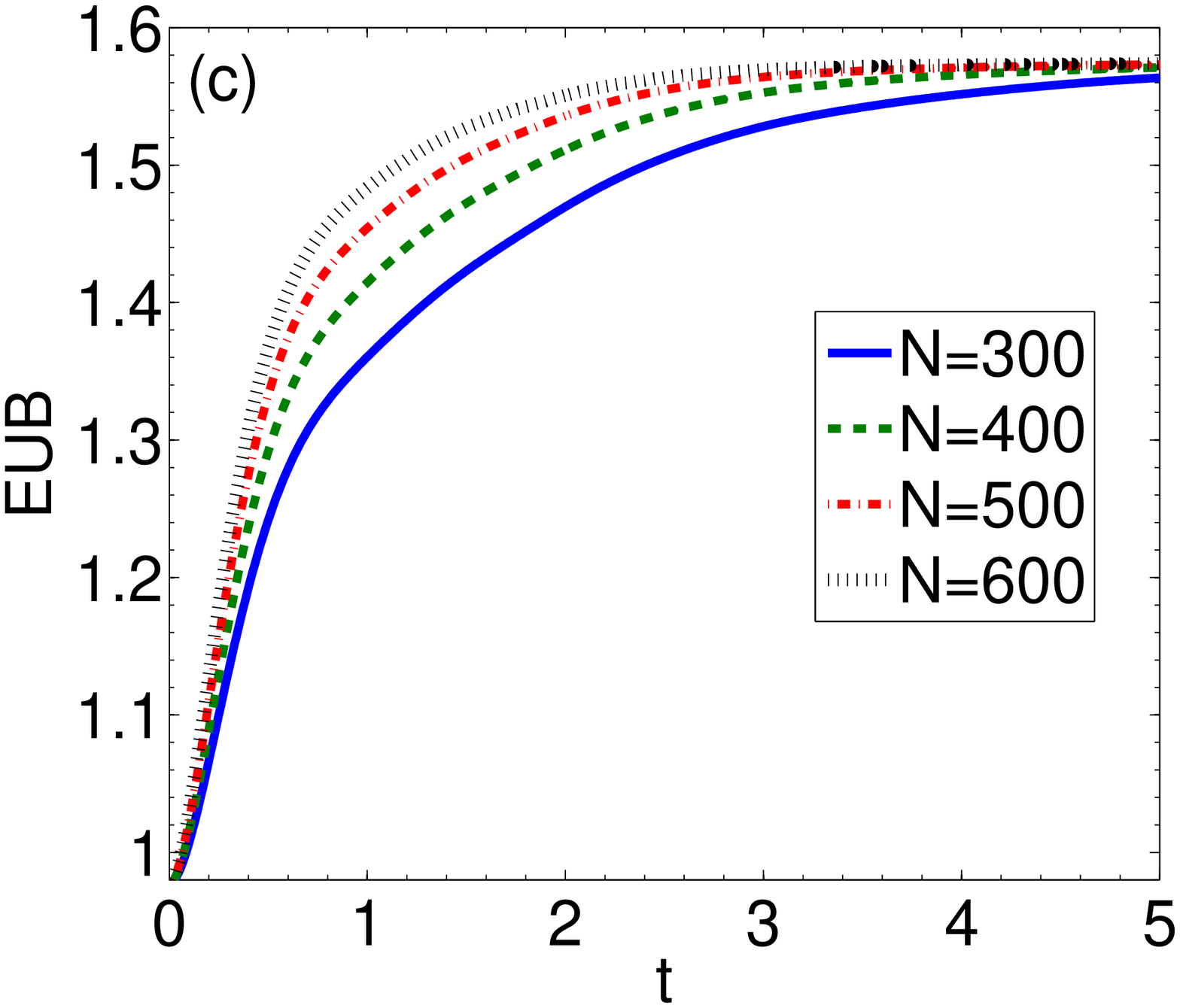}
  \includegraphics[width=0.49\textwidth]{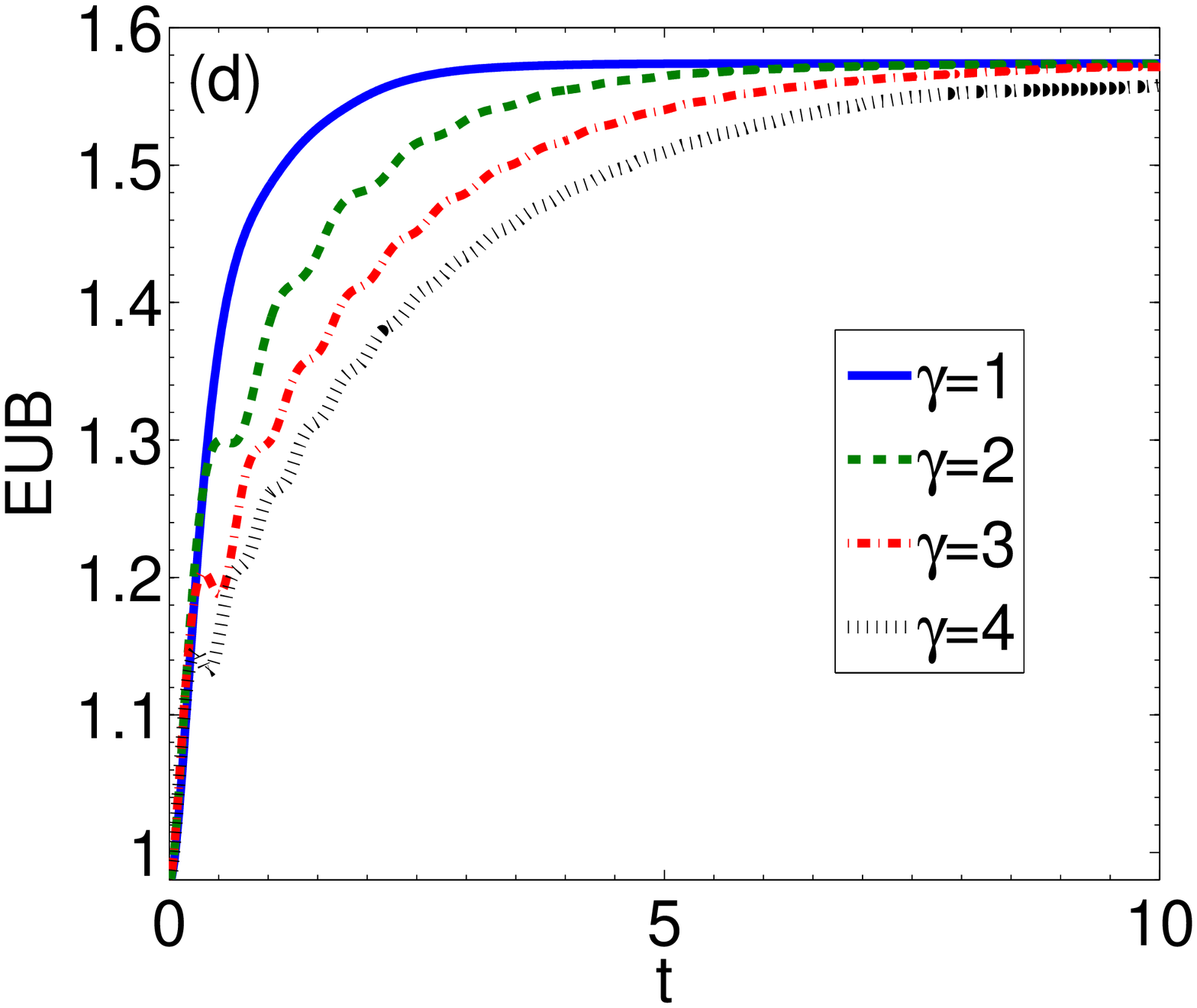}
\caption{The dynamics of entropic uncertainty bound for the case in which the initial state is mixed with $r_1=1$, $r_2=-0.2$, and $r_3=0.2$. \textbf{(a)} For different values of the strength of transverse magnetic field $\lambda$, when $D=0$, $\gamma=1$, $\delta=0$, $g=0.05$ and $N=600$. \textbf{(b)} For different values of DM interaction $D$, when $\lambda=1$, $\gamma=1$, $\delta=0$, $g=0.05$ and $N=600$. \textbf{(c)} For different values of the spin number $N$, when $\lambda=1$, $\gamma=1$, $\delta=0$, $g=0.05$ and $D=0$. \textbf{(d)} For different values of the anisotropy $\gamma$, when $D=0$, $\lambda=1$, $\delta=0$, $g=0.05$ and $N=600$.}
\label{figure2}
\end{figure}

The dynamics of EUB for different values of the number of the total sites in XY spin chain environment $N$ is sketched in Fig. \ref{figure1}(c). It is observed that with increasing the number of sites in the spin chain, the environment has a stronger effect on the system and more information is lost about the system. Hence, the uncertainty will be increased by the number of sites $N$ in the spin chain environment.
In Fig. \ref{figure1}(d), the effect of the anisotropy of exchange interaction on the dynamics of EUB is investigated. By evaluating this plot, it becomes clear that the EUB can be suppressed by increasing the anisotropy parameter $\gamma$.
In Fig. \ref{figure2}, the EUB (\ref{eub1}) is plotted in terms of time for the case in which the initial state that is shared between Alice and Bob is mixed with the state parameters $r_1=1$, $r_2=-0.2$, and $r_3=0.2$, i.e. the initial state becomes the mixed state as $\rho_{AB}(0)=0.6|\Phi\rangle\langle\Phi|+0.4|\Psi\rangle\langle\Psi|$ where $\Phi=(|00\rangle+|11\rangle)/\sqrt{2}$ and $\Psi=(|01\rangle+|10\rangle)/\sqrt{2}$ are Bell states. The qualitative behavior of uncertainty in this case is similar to that of pure state case. In terms of quantitative description of uncertainty, we must mention that the EUB at the initial moment of interaction tends to one. Of course, it is worth noting that when the interaction begins, it is observed that the uncertainty bound increases with the continuation of the interaction.

Fig. \ref{figure2}(a) represents the dynamics of EUB for different values of strength of transverse magnetic field $\lambda$. It is observed that the EUB decreases with increasing parameter $\lambda$. In fact, increasing $\lambda$  suppresses the EUB.
Fig. \ref{figure2}(b) displays the time evolution of EUB for four values of DM interaction strength with fixed values of $\lambda=\gamma=1, \delta=0, g=0.05,$ and $N=600$.  From this plot, one concludes that the EUB is enhanced by increasing the strength of DM interaction, which behaves similarly to the previous case (pure state case).

The time evolution of EUB for different values of the number of the total sites in XY spin chain environment $N$ is depicted in Fig. \ref{figure2}(c). It is observed that with increasing the number of sites in the spin chain, the environment has a stronger effect on the system and more information is lost about the system. So, as mentioned before, the uncertainty will be enhanced by the number of sites $N$ in the spin chain environment.
In Fig. \ref{figure2}(d), the effect of the anisotropy of exchange interaction on the dynamics of EUB is analyzed. As can be seen, the EUB can be repressed by increasing the anisotropy parameter $\gamma$.

\section{Conclusions}\label{sec4}
To summarize, we have studied the time evolution of EUB for a two-qubit state coupled to a spin chain with DM interaction. In the study of quantum systems, it must be borne in mind that the interaction of the system with its surroundings in the real world is inevitable. As a result of the interaction of the system with the environment, the information is lost from the quantum system. Specifically, we considered a model in which the bipartite quantum system is available to Alice and Bob so that parts $A$ and $B$ are owned by Alice and Bob, respectively. We also assumed that the bipartite quantum system which is coupled to a spin chain with DM interaction. As a result of the interaction between the system and an environment, Bob's information about Alice's system decreases, and vice versa. Therefore, as expected, Bob's uncertainty about the result of Alice's measurement on her system increases. In this work, we examined the effect of environmental parameters on EUB. Notably, it was observed that the EUB can be suppressed by raising the strength of the transverse magnetic field $\lambda$ and anisotropy of exchange interaction $\gamma$.  On the other side, it was also seen that the EUB can be enhanced by increasing the number of sites $N$ and the strength of DM interaction. To sum up, our research may open a new window on the time evolution of EUB in the Heisenberg spin chain models and be of interest to quantum measurement precision in quantum information processing.

\section*{ORCID iDs}
Soroush Haseli \href{https://orcid.org/0000-0003-1031-4815}{https://orcid.org/0000-0003-1031-4815}\\
Saeed Haddadi \href{https://orcid.org/0000-0002-1596-0763}{https://orcid.org/0000-0002-1596-0763}\\
Mohammad Reza Pourkarimi \href{https://orcid.org/0000-0002-8554-1396}{https://orcid.org/0000-0002-8554-1396}\\

\end{document}